\documentclass[runningheads]{llncs}
\usepackage{graphicx}
\usepackage{tabularx}
\usepackage[inline,shortlabels]{enumitem}
\usepackage{verbatim}
\usepackage{amsmath}
\usepackage {xcolor}
\usepackage{commath}
\usepackage{comment}
\usepackage{amsmath}
\usepackage{hyperref}
\usepackage[misc,geometry]{ifsym}
\begin{document}
\title{Patient-independent Schizophrenia Relapse Prediction Using Mobile Sensor based Daily Behavioral Rhythm Changes}
\titlerunning{Schizophrenia Relapse Prediction Using Mobile Sensing Behavioral Rhythms}
%

\author{{Bishal Lamichhane \inst{1}}\Letter \and
Dror Ben-Zeev\inst{2} \and
Andrew Campbell\inst{3} \and
Tanzeem Choudhury\inst{4} \and
Marta Hauser\inst{5} \and
John Kane\inst{5} \and
Mikio Obuchi\inst{3} \and
Emily Scherer\inst{3} \and
Megan Walsh\inst{5} \and
Rui Wang\inst{3} \and
Weichen Wang\inst{3} \and
Akane Sano\inst{1}}

\authorrunning{B. Lamichhane et al.}
%
\institute{Rice University, USA \\
\email{bishal.lamichhane@rice.edu} \\ \and
University of Washington, USA \and
Dartmouth College, USA \and
Cornell University, USA \and
Northwell Health, USA}

\maketitle              
\begin{abstract}

A schizophrenia relapse has severe consequences for a patient’s health, work, and sometimes even life safety. If an oncoming relapse can be predicted on time, for example by detecting early behavioral changes in patients, then interventions could be provided to prevent the relapse. In this work, we investigated a machine learning based schizophrenia relapse prediction model using mobile sensing data to characterize behavioral features. A patient-independent model providing sequential predictions, closely representing the clinical deployment scenario for relapse prediction, was evaluated. The model uses the mobile sensing data from the recent four weeks to predict an oncoming relapse in the next week. We used the behavioral rhythm features extracted from daily templates of mobile sensing data, self-reported symptoms collected via EMA (Ecological Momentary Assessment), and demographics to compare different classifiers for the relapse prediction. Naive Bayes based model gave the best results with an F2 score of 0.083 when evaluated in a dataset consisting of 63 schizophrenia patients, each monitored for up to a year. The obtained F2 score, though low, is better than the baseline performance of random classification (F2 score of 0.02  $\pm$ 0.024). Thus, mobile sensing has predictive value for detecting an oncoming relapse and needs further investigation to improve the current performance. Towards that end, further feature engineering and model personalization based on the behavioral idiosyncrasies of a patient could be helpful.

\end{abstract}

\section{Introduction}
\label{Introduction}

Schizophrenia is a chronic mental disorder affecting about 20 million people worldwide \cite{James2018}. Patients with schizophrenia perceive reality abnormally and show disturbances in their thoughts and behaviors. Some of the associated symptoms are delusions, hallucinations, disordered thinking, incoherent speech (putting together words that do not make sense), disorganized motor functions, social withdrawal, appearances of lack of emotions, etc. \cite{Andreasen1991,Jablensky2010}. A patient with schizophrenia is generally treated with antipsychotic drugs and psycho-social counseling. These patients are treated as out-patients, in the general non-serious cases, and they visit the clinic for routine mental health assessment. During the visit, the patient's symptoms are tracked and medications/therapies are adapted. Questionnaires such as BPRS (Brief Psychiatric Rating Scale) \cite{Overall1962} are used to keep track of the symptoms. A patient with schizophrenia under a treatment regimen might sometimes experience a relapse, an acute increase of schizophrenia symptoms and degrading mental health. The routine clinical visits and BPRS based assessments are meant to keep track of symptoms and prevent any likely relapses. However, the clinical visits happen only every few months and a patient might have a relapse in between the visits.

A relapse has severe consequences for both the patients and their caregivers (e.g. their family), even endangering their lives in some cases. So it is important to detect an oncoming relapse and provide timely interventions for prevention. It might be possible to use mobile sensing to predict an oncoming relapse by detecting behavioral and emotional changes associated with schizophrenia symptoms. Mobile sensors like accelerometer, GPS, ambient light sensors, microphones, etc. can capture various aspects of a person's behavior. These can then be complemented by questionnaires (e.g. Ecological Momentary Assessments - EMA), delivered through a mobile application, to assess the person's self-reported symptoms, behavior and feeling and build a relapse prediction model. Mobile sensing would be a low-cost and scalable solution for relapse prediction compared to other alternatives such as the pharmacological approach \cite{Lieberman1987}.

In this work, we investigated mobile sensing based schizophrenia relapse prediction using mobile sensing. Relapse prediction is framed as a binary classification problem, associating an upcoming period as relapse or non-relapse based on the features observed in the current period. We extracted daily behavioral rhythm based features from mobile sensing data, which was also effective in predicting self-reported schizophrenia symptoms in our previous work \cite{Tseng2020}, complemented by self-reported symptoms collected through EMA and demographics features, and evaluated different classifiers for relapse prediction. Daily template based rhythm features were found to outperform feature sets proposed in previous works for relapse prediction. Further, our proposed model is a sequential prediction model trained and evaluated in a patient-independent setting. Such a relapse prediction model, closer to a clinical deployment solution, has not been investigated in previous works. Our work establishes the basic feasibility of using mobile sensing for schizophrenia relapse prediction and identifies related challenges, to be addressed in future work. The paper is organized as follows. In Section \ref{section:previous_work}, we present some of the related works on relapse prediction in the context of schizophrenia and other mental disorders. In Section \ref{section:methods}, the dataset and methodology used for developing the relapse prediction model are discussed. This is followed by Section \ref{section:results} where we present the evaluation results of the developed model. These results are discussed in Section \ref{section:discussion} and we present our conclusions in Section \ref{section:conclusion}.

\section{Related Work}
\label{section:previous_work}

Several previous works have investigated the prediction of relapses in the context of mental disorders and substance addiction. The authors in \cite{Birnbaum2019} studied the prediction of psychotic symptoms relapses based on the linguistic and behavioral features inferred from the Facebook post. The prediction model was evaluated to have a specificity of 0.71 in a study of 51 participants. The work thus showcased the potential of behavior profiling for relapse prediction in the context of mental disorders. In \cite{Matcham2019}, the authors are aiming to use mobile sensing based features such as sleep quality, sociability, mobility, and mood changes to predict the relapse of depressive episodes. Mobile sensing and social behavior (online or offline social behavior) have also been found to be helpful in predicting relapses of substance addictions. The authors in \cite{yang2018} analyzed social media posts and social network influences to predict the relapse of opioid usages. Similarly, the authors in \cite{Bishop2020} discussed the relevance of several contextual information such as sleep deprivation, affect, environment, and location, derivable from mobile sensing, for predicting relapse of alcoholism.

Some earlier works have already investigated schizophrenia relapse prediction based on mobile sensing. For example, the authors in \cite{Barnett2018} investigated the relation of schizophrenia relapse with mobility and behavioral features derived from mobile sensing. In their study population of 17 patients, 5 patients had a relapse. The authors analyzed the anomaly of mobility and sociability features in this population and found increased incidences of an anomaly in weeks leading up to relapse. The anomaly was defined as the deviation of features from an expected pattern. Though this work is one of the pioneering works on mobile sensing based schizophrenia relapse prediction, generating novel qualitative insights, the authors did not develop any prediction model probably due to the limited size of the study population. The authors in \cite{Buck2019} also explored the usage of mobile sensing based features for schizophrenia relapse prediction. Sociability features based on outgoing calls and messages were found to be significantly different before a relapse, compared to a non-relapse period. This insight is helpful to predict an oncoming relapse. However, the others only provided qualitative analysis and no predictive models were evaluated. In contrast to these two earlier works which offered qualitative analysis only, we proposed and evaluated relapse prediction models in our work.

Relapse prediction models have been investigated in a previous work of \cite{Wang2018,Wang2020}. The authors evaluated the potential of mobile sensing based features to predict an oncoming relapse. The authors also framed relapse prediction as a binary classification task, classifying an oncoming period as either relapse or non-relapse. Mobility, sociability, and EMA features were computed for each epoch of the day (morning, afternoon, evening, and night) and features from N days (comparing for different values of N) were used to predict if there was going to be a relapse in the next day. Several machine learning models were evaluated for relapse prediction. Using 3-fold cross-validation, SVM (rbf kernel) was found to give the best performance with an F1-score of 0.27. The study population consisted of 61 patients with schizophrenia where 27 instances of relapse were reported in 20 patients. We used the same dataset for our evaluations and build upon the work of the authors to generate further insights on a mobile sensing based relapse prediction model. The authors in \cite{Wang2020} established that the mobile sensing based behavioral features indeed have an association with an upcoming relapse. However, there was likely a look-ahead bias in their evaluations due to k-fold random cross-validation that was used. Within k-fold cross-validation, mobile sensing data from the future is also used for building a prediction model for a given test patient, while the model is being evaluated using the currently observed data. In contrast to this approach, we developed a sequential relapse prediction model evaluated in a patient-independent setting. The relation between current/past mobile sensing data and future relapses is first modeled from the patients in the training set only. The trained model is then used to predict, sequentially over time, if the mobile sensing data from the patient in the test set indicate an oncoming relapse. This approach of modeling brings the evaluation closer to clinical deployment. Further, unlike the work in \cite{Wang2020}, we do not impose any knowledge of relapse location to create the feature extraction/evaluation windows. Its implication is that a sliding window approach to relapse prediction has to be used, leading to a higher number of feature extraction windows to be evaluated. Naturally, this leads to a higher chance of incurring false positives during prediction and reduced classification performance. Nonetheless, such an evaluation would better reflect a real clinical deployment scenario. Finally, we used the daily behavioral rhythm features extracted from the daily template, composed of the hourly averages of the mobile sensing data, to characterize the behavioral patterns of a patient. Thus, finer temporal resolution is retained for feature extraction compared to the work of \cite{Wang2020}  where features were extracted for each of the 6-hour periods of the day (6 am - 12 pm, 12 pm - 6 pm, 6 pm - 12 am, and 12 am - 6 am).    

\section{Methods}
\label{section:methods}

In this section, we describe the dataset and methodology that has been used to develop our proposed relapse prediction model.

\subsection{Dataset}
\label{Dataset}

We used the dataset from the CrossCheck project \cite{Wang2016,Wang2017,Wang2018,Buck2019} (available at \href{https://www.kaggle.com/dartweichen/crosscheck}{https://www.kaggle.com/dartweichen/crosscheck}) for the development and evaluation of a relapse prediction model. The dataset consists of data from a clinical trial where 75 schizophrenia patients were monitored for up to a year with the Crosscheck system \cite{BenZeev2017} continuously collecting passive sensing data from patients' smartphones. The number of patients and the monitoring period are significantly larger than those in previous works on schizophrenia patient monitoring \cite{Barnett2018}. The data collected were: accelerometer, light levels, sound levels, GPS, and call/sms log. Further, the Crosscheck system also routinely obtained self-assessments from patients with EMA (Ecological Momentary Assessment) \cite{Shiffman2008}. These EMA, which were obtained up to three times in a week, consisted of 10 questions to assess patients' current emotional and behavioral state. The questions asked were, for instance, \textit{Have you been feeling calm?, Have you been social?} etc. Patients could answer the EMA questions with four options: \textit{Not at all, A little, Moderately, Extremely}. EMA obtained at a low frequency, e.g. every few days only, makes it less burdensome for the patients. In the dataset, data from 63 patients were made available for analysis. The mean age and education years of these patients were: 37.2 years (min:18 years, max: 65 years) and 9.4 years (min:5 years, max:14 years) respectively. Among the 63 patients, 20 patients had a relapse and there were 27 instances of relapse in total (some patients had multiple relapses) as annotated by clinical assessors \cite{Wang2018,Wang2020}.

\subsection{Relapse Prediction Model}
\label{section:relapse_prediction_model}
\begin{figure}[!htb]
\centering
\includegraphics[scale=0.35]{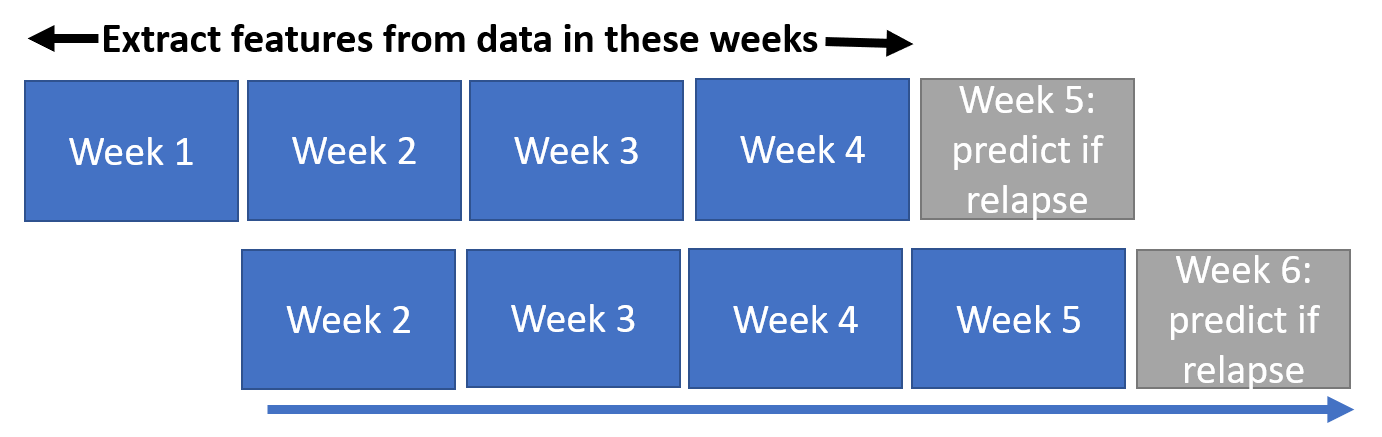}
\caption{\label{fig:relapse_prediction_window}Relapse prediction approach in our model. Features are extracted from 4 weeks of data to predict an oncoming relapse in the next week. A prediction for each week is produced with a sliding window of stride length 1 week.}
\end{figure}

We developed machine learning models that can predict if there is an oncoming relapse in the next week (prediction window) based on the mobile sensing data from recent 4 weeks (feature extraction window). A sliding window with a stride of 1 week is used for feature extraction, thus obtaining a sequential prediction for each week of monitoring. This approach of relapse prediction is shown in Figure \ref{fig:relapse_prediction_window}. We trained and evaluated the model in a patient-independent setting, using leave-one-patient-out cross-validation. The features that were used for our relapse prediction model are described next.

\subsection{Features}
\label{section:features}

A summary of all the features extracted from the daily template (composed of hourly averages) of mobile sensing data, EMA, and demographics data are shown in Table \ref{Table:features_summary}. In this section, we describe how these different features are extracted.

\subsubsection{Daily rhythm features}

\begin{figure}[!htb]
\centering
\includegraphics[width=\columnwidth]{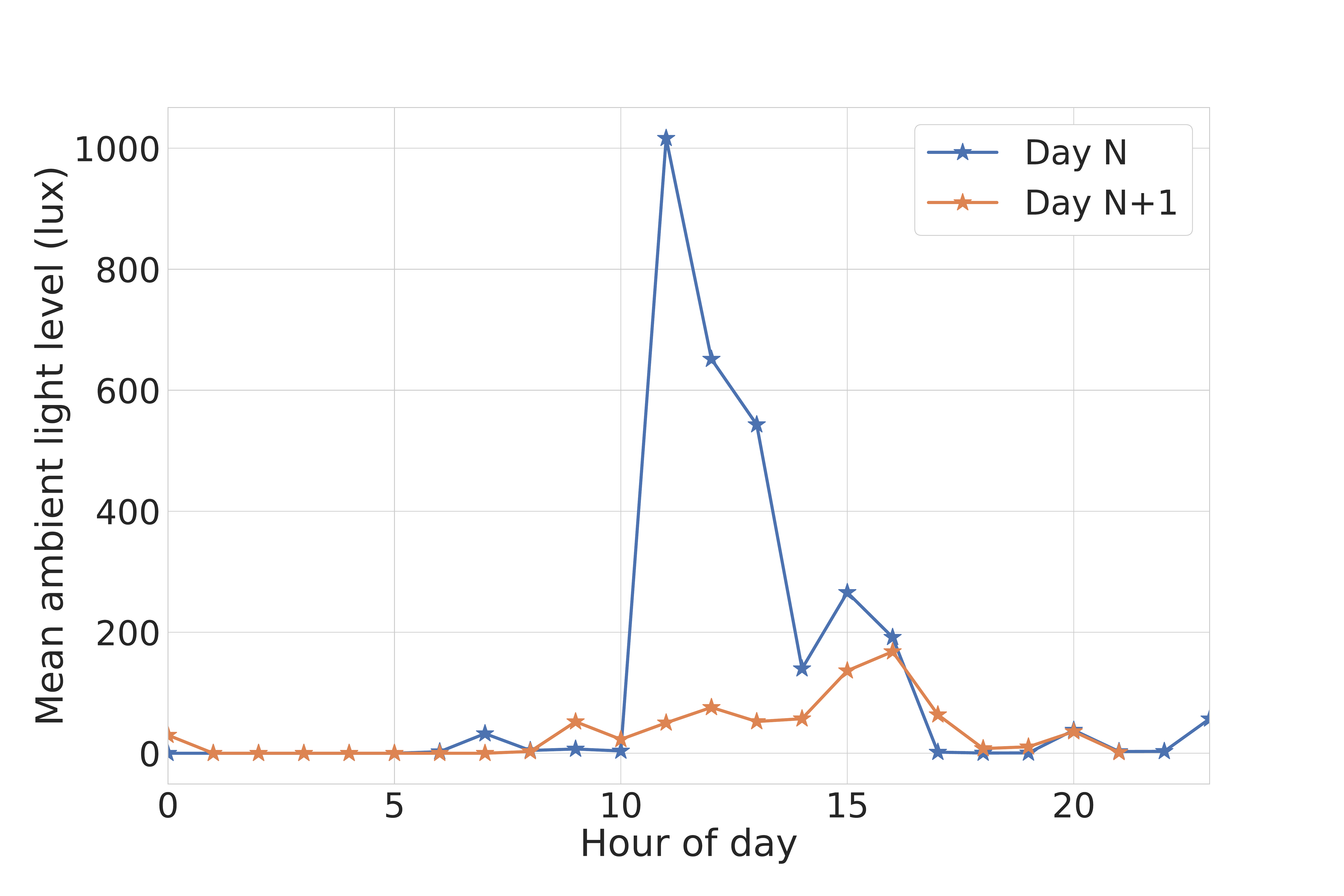}
\caption{\label{fig:light_daily_template} An example showing the daily template for the ambient light levels in two consecutive days, obtained from the hourly averages of the light levels recorded from the smartphone of a patient for the given days. Daily templates were obtained for six different signal modalities available: accelerometer magnitude, light levels, distance, conversation, sound levels, and call duration. These templates were then characterized to obtain daily template features used for relapse prediction.}
\end{figure}

Six mobile sensing signals, obtained continuously throughout the day, were derived from the dataset for daily template based rhythm feature extraction. The signals derived were: accelerometer magnitude (magnitude from 3-axis accelerometer signal recordings), (ambient) light levels, distance traveled (from GPS), call duration (from call log), sound levels, and conversation duration. These signals were derived from the raw mobile sensor recordings as in \cite{Wang2020}. We obtained a daily template for each of the mobile sensing signals by computing the hourly averages of the signal in a given day of monitoring (thus the template consists of 24 points corresponding to each hour of the day). The templates capture daily rhythmic behaviors which are relevant for monitoring behavioral changes in schizophrenia patients\cite{Tseng2020}. An example of a daily template obtained for the light level signal is shown in Figure \ref{fig:light_daily_template}. Five categories of features were extracted from the daily templates of mobile sensing signals.

\begin{enumerate}[(i)]
    
\item Mean daily template features: Since we used a feature extraction window of 4 weeks, there are 28 daily templates of each of the mobile sensing signals in a given feature extraction window. The daily templates of a mobile sensing signal across the 4 weeks were averaged to obtain the mean daily template (mDT). An example is shown in Figure \ref{fig:mean_light_daily_template}. The obtained mDT was then characterized by six statistical features: mean, maximum, standard deviation, range, skewness, and kurtosis.

\begin{figure}[!htb]
\centering
\includegraphics[width=\columnwidth]{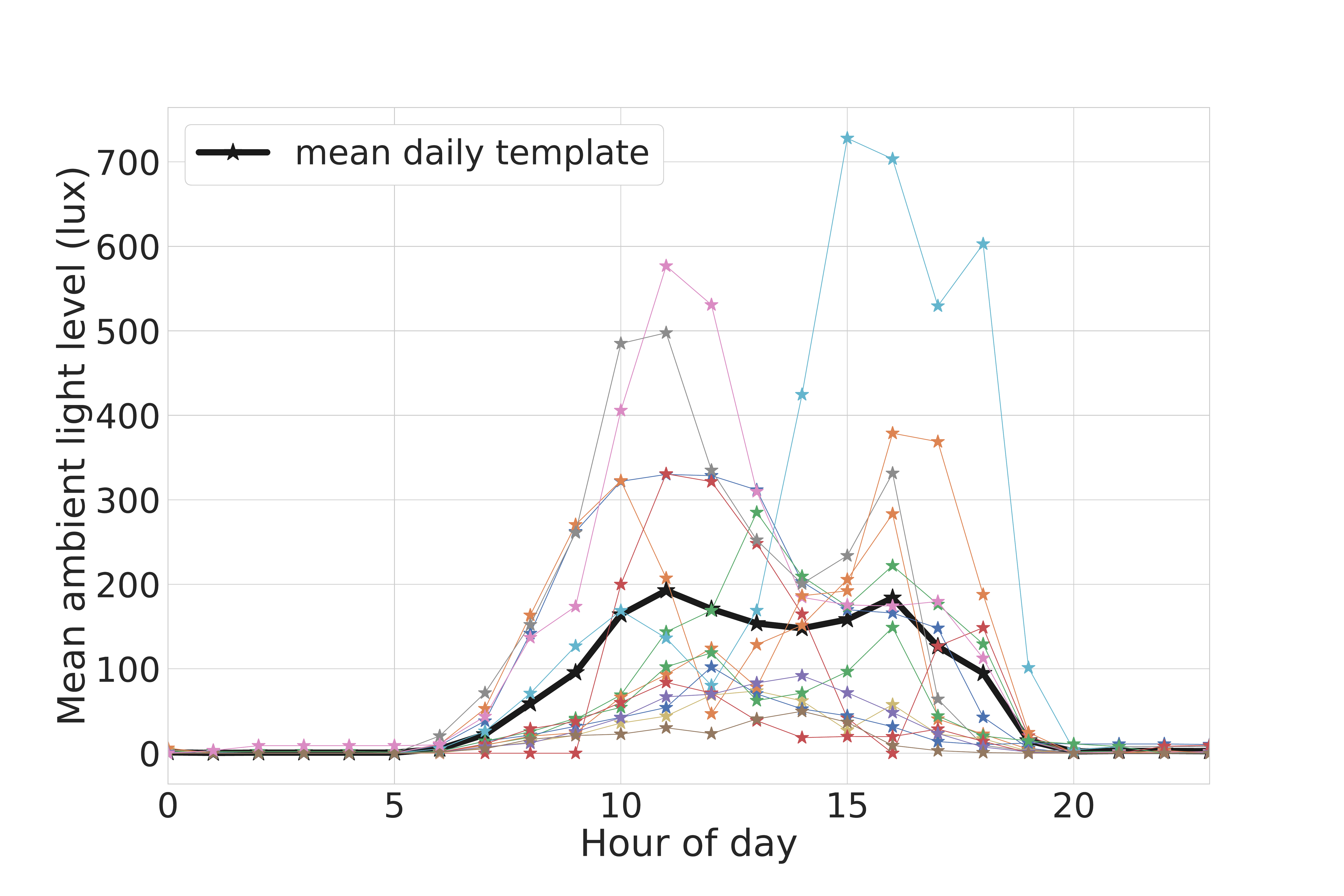}
\caption{\label{fig:mean_light_daily_template} Mean daily template obtained by averaging the daily templates of the days within a feature extraction window.}
\end{figure}

\item Deviation daily template features: Just like the mean daily template which was obtained by averaging the 28 daily templates in a feature extraction window, deviation daily template (dDT) was obtained by taking the standard deviation of the daily templates (deviation of each of the points in the template) across the 28 days, for each patient. The mean of the obtained dDT was then extracted as a feature to characterize the signal variability in a given feature extraction window. 

\item Maximum daily template features: Maximum daily template (mxDT) was obtained similarly as mDT and dDT by taking the maximum of the hourly average points across 28 days in the daily templates, within a feature extraction window. For computing features from mxDT, the difference between mDT and mxDT was obtained and the maximum absolute difference ($\text{maxDiff} = max(|mDT-mxDT|)$) was extracted as a feature characterizing the maximum deviation from the mean in a given feature extraction window. 

\item Template distance features: To characterize the changes between successive feature extraction windows, we computed features based on the distance between the templates for the current and the previous feature extraction window. In particular, distance based on mDT and mxDT were used. First, we normalized mDT and mxDT of a feature extraction window with their respective maximum value. Then the distance between the normalized mDTs (mDT for the current and the previous feature extraction window) was computed as a feature with:
\begin{equation*}
    dist_{mDT} = \sum_{i=1}^{24} ((mDT_{norm}(curr)[i] - mDT_{norm}(prev)[i])^2) 
\end{equation*}

A weighted version of $dist_{mDT}$, considering the points in the template between 9 AM - 9 PM only, was also extracted as a feature to characterize the differences seen in the main part of the day. The weighted distance was computed as:
\begin{equation*}
wdist_{mDT} = \sum_{i=9}^{21} ((mDT_{norm}(curr)[i] - mDT_{norm}(prev)[i])^2) 
\end{equation*}

The mxDT based distance feature was computed between the normalized mxDT and mDT as:
\begin{equation*}
dist_{mxDT} = \sum_{i=1}^{24}((mxDT_{norm}(curr)[i] - mDT_{norm}(prev)[i])^2) 
\end{equation*}

\item Signal mean and variability: The daily template (Figure \ref{fig:light_daily_template}) consists of hourly averaged values of mobile sensing signal modalities for each day. Daily averages for each mobile sensing signal can be estimated by taking the average of the points in the daily template. From these daily averages in a given feature extraction window, mean and variability (standard deviation) of a mobile sensing signal were computed as features.

\end{enumerate}

\subsubsection{EMA and Demographics features}

Besides the daily template-based behavioral rhythm features extracted from the mobile sensing signals, we also computed features from the 10-item EMA data (Section \ref{Dataset}) in a given feature extraction window. For each of the EMA items, we computed its mean and standard deviation within the window as features. Thus a total of 20 features are extracted from the EMA data. Behavioral features and relapse characteristics might also be dependent on the demographics (e.g. age group of a patient). To allow for the implicit personalization of the relapse prediction model, we included the age of the patient and their year of education (which could be a surrogate for their work type) as demographic features. These demographic features (dimension 2) were appended alongside the EMA  features (dimension 20) and daily template features (dimension 78) for each of the feature extraction window to characterize the behavioral patterns in a given window.

\begin{table}[!htb]
\caption{\label{Table:features_summary} Different features extracted from the mobile sensing, EMA, and demographics for relapse prediction. Features are extracted from six mobile sensing signal using their daily template representation and 10 items of the EMA.} 
\begin{tabularx}{\linewidth}{X}
\hline 
\textbf{Daily Rhythm features} \\ 
\hline 
Mean daily template (mDT) features: \textit{ mean, standard deviation, maximum,
range, skewness, kurtosis} \\ 
Standard deviation template (sDT) features: \textit{mean} \\
Absolute difference between mDT and mxDT: \textit{maximum} \\
Distance between normalized mDT(current) and mDT(previous) \\
Weighted distance between normalized mDT(current) and mDT(previous) \\
Distance between normlized mxDT(current) and mDT(previous)\\
Daily averages: \textit{mean, standard deviation} \\
\hline 
\textbf{EMA features} \\
\hline 
EMA item values: \textit{mean, standard deviation} \\
\hline
\textbf{Demographics}\\
\hline
Age, Education years \\
\hline
\end{tabularx}
\end{table}

\subsection{Classification}

\subsubsection{Dataset size}
With our feature extraction and prediction window sizes (Section \ref{section:relapse_prediction_model}), we obtained a total of 2386 feature extraction windows from the entire dataset. Of these, 23 windows were labeled as preceding (by a week) an incidence of relapse. Some of the relapse incidents got excluded from the analysis as they were too early in the monitoring period or there was no monitoring data around the relapse dates. When a feature extraction window was identified as preceding a relapse, then the next feature extraction window was obtained after a cool-off period of 28 days (similar to the cool-off period concept used in \cite{Wang2020}). This was done to prevent any feature extraction window from being corrupted by monitoring data during the actual relapse which might include hospitalization or other interventions. 

\subsubsection{Model validation}
We used leave-one-patient-out cross-validation for the validation of the relapse prediction model. Data from all the patients, except from one (hold-out set), was used to train a classifier for relapse prediction. The trained model was then evaluated using the data from the hold-out patient. This process was repeated with a different patient in the hold out set every time. Leave-one-patient-out for model validation reflects a clinical deployment scenario where a trained model is expected to provide predictions for a new unseen patient. The trained model could be adapted for the new patient with different model personalization strategies.   

\subsubsection{Classifiers}
As our dataset size is fairly small, a simple classifier could be more suited for the classification task. Therefore we chose to evaluate Naive Bayes based classification for relapse prediction. We also evaluated other classifiers for comparison. In particular, we evaluated Balanced Random Forest (BRF) \cite{Chao2004} and EasyEnsemble (EE) classifier \cite{Liu2009}. These classifiers were selected since they are suited for learning in an imbalanced dataset (The ratio of relapse to non-relapse is $\sim$1:100 in our classification task and is thus imbalanced). Isolation Forest (IF) \cite{Liu2008}, a one-class classifier commonly used for outlier detection, was also evaluated. In the IF based classification, the relapse class was treated as the outlier class. The number of trees for BRF, EE, and IF was empirically set to 51, 101, and 101 respectively. We also evaluated a classification baseline by randomly predicting relapse or non-relapse for each week of prediction in the test set (within the leave-one-patient-out cross-validation setting). The ratio of relapse to non-relapse in these random predictions was matched to the ratio in the training set. The random predictions for 1000 independent runs were averaged to obtain the baseline results. 

\textit{Feature transformation:} There are different flavors of Naive Bayes classifier, each imposing an assumption on the distribution of the underlying features. We used the Categorical Naive Bayes model since features can be easily transformed to be categorical with simple transformations. We transformed each of the features extracted (Section \ref{section:features}) into 15 categories (empirically chosen) based on the bin membership of each feature values in its histogram. The histogram is constructed from the training data only. These transformed features were then used in a categorical Naive Bayes classification model. The categorization of features quantizes the behavioral patterns and relapses could be linked as a shift in the categorized levels. Feature transformation with categorization was found to be beneficial (better classification performance) for use with the other classifiers considered (BRF, EE, and IF). Thus we employed feature transformation in the classification pipeline irrespective of the classifier used.

\textit{Feature selection:}
Since we extracted a large number of features and our dataset size is relatively small, we evaluated the classification pipeline with a patient-specific feature selection strategy. A training sub-sample, consisting of all the data points labeled as relapse in the training set and N non-relapse data points from the training set patients closest in age to the patient in the test set, is selected. From this training sub-sample, M top features are identified. We used mutual information based criterion between features and the target label to select the top M features. The machine learning model for relapse prediction was then trained using these selected M features only. In our leave-one-patient-out cross-validation, different feature sets would be automatically selected depending upon the patient currently in the test set. The value of N was set to 100 (so that non-relapse data from at least two patients are included in the training subset) and M was set to 5 (which gave the best performance from the considered values: 3, 5, 10, 15, and 20). The underlying hypothesis for the age-based training sub-sample creation is that the patients from a similar age group would have similarities in their behavior and thus the feature-target relations would translate within the age groups.

\subsection{Evaluation Metric}

We evaluated the relapse prediction model using F2 score metric which is defined as:
\begin{align*}
F2 = \frac{5*precision*recall}{4*precision+recall}
\end{align*}
where $precision = \frac{TP}{TP+FP}$ and $recall=\frac{TP}{TP+FN}$ (TP: Number of True positives, FP: Number of False positives, FN: Number of False Negatives). F2 score gives higher priority to recall compared to precision. In the context of the relapse prediction task, this translates to higher importance assigned for correctly predicting an oncoming relapse which is more important than the associated trade-off of avoiding a false alarm.

\section{Results}
\label{section:results}

\subsubsection{Classifier comparison}
We evaluated the classification performance with different machine learning models using leave-one-patient-out cross-validation. The obtained results are given in Table~\ref{Table:results_comparison}. Naive Bayes based classification gives the best classification performance with an F2 score of 0.083.

\begin{table}[!htb]
\centering
\caption{\label{Table:results_comparison} Comparison of different classifiers for relapse prediction models. Features from the daily template of mobile signal data, EMA, and demographics are used for the classifier.}{
\begin{tabular}{|c|c|c|c|}
\hline 
\textbf{Method} & \textbf{F2-score} & \textbf{precision} & \textbf{recall} \\ 
\hline 
Naive Bayes & \textbf{0.083} & 0.22 & 0.086\\
\hline 
Balanced Random Forest  & 0.042 & 0.01 & 0.47 \\
\hline 
EasyEnsemble & 0.034 & 0.007 & 0.43 \\ 
\hline
Isolation Forests & 0.045 & 0.01 & 0.39 \\
\hline 
Random classification baseline & 0.020 $\pm$ 0.024 & 0.010 $\pm$ 0.012 & 0.026  $\pm$ 0.032\\
\hline

\end{tabular}} 
\end{table}

\subsubsection{Feature comparison}
In our work, we computed daily template based rhythm features to characterize behavioral patterns and changes. We compared the classification performance obtained with this feature set to that obtained using the feature set from \cite{Wang2020} where features are computed with lower temporal resolution. To optimize the classification pipeline using the feature set from \cite{Wang2020}, we selected the best parameter (training subset size for feature selection N, and the number of selected features M) using grid search. Similarly, the demographic features were also added as it improved the classification performance. The obtained results using Naive Bayes model, which provided the best performance in both of the feature sets, are given in Table \ref{Table:feature_comparison}.

\begin{table}[!htb]
\centering
\caption{\label{Table:feature_comparison} Comparison of feature sets for the relapse prediction task. The features based on the daily templates, where the hourly averages of the mobile sensing signal are retained, are compared with the features from \cite{Wang2020} where features are computed with a lower temporal resolution (6 hours).}{
\begin{tabular}{|c|c|}
\hline 
\textbf{Feature set} & \textbf{F2-score} \\ 
\hline 
Daily template based, EMA, demographics (this work) & \textbf{0.083} \\
\hline 
Feature set from \cite{Wang2020} & 0.065 \\
\hline

\end{tabular}} 
\end{table}

\subsubsection{Modality comparison}
\begin{figure}[!htb]
\centering
\includegraphics[width=\columnwidth]{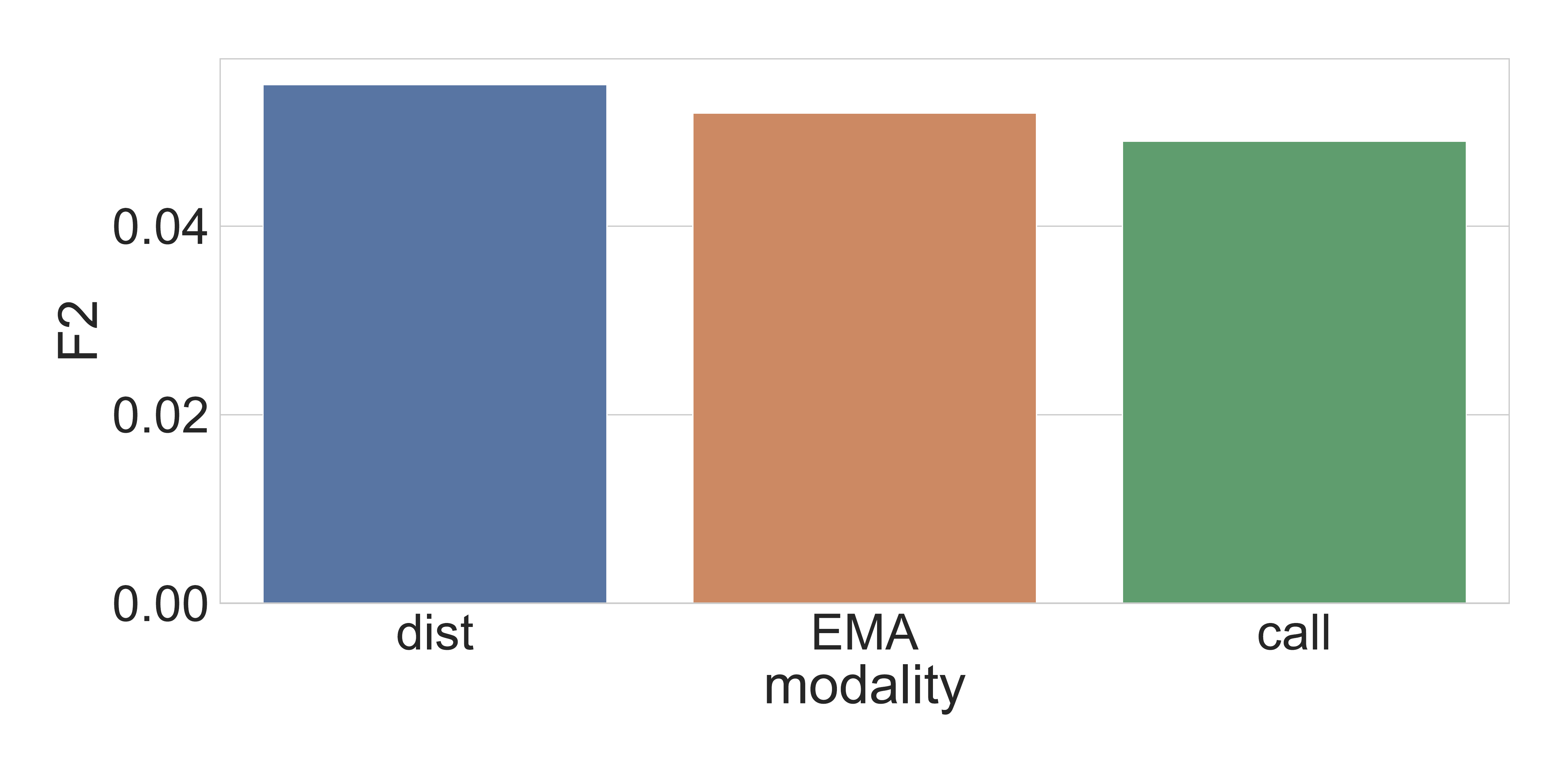}
\caption{\label{fig:per_modality_f2} F2 score obtained with different signal modalities (top 3 modalities) for the relapse prediction task. Distance traveled (dist) is found to be most relevant for relapse prediction followed by EMA and call duration (call).}
\end{figure}

\begin{figure}[!htb]
\centering
\includegraphics[width=\columnwidth]{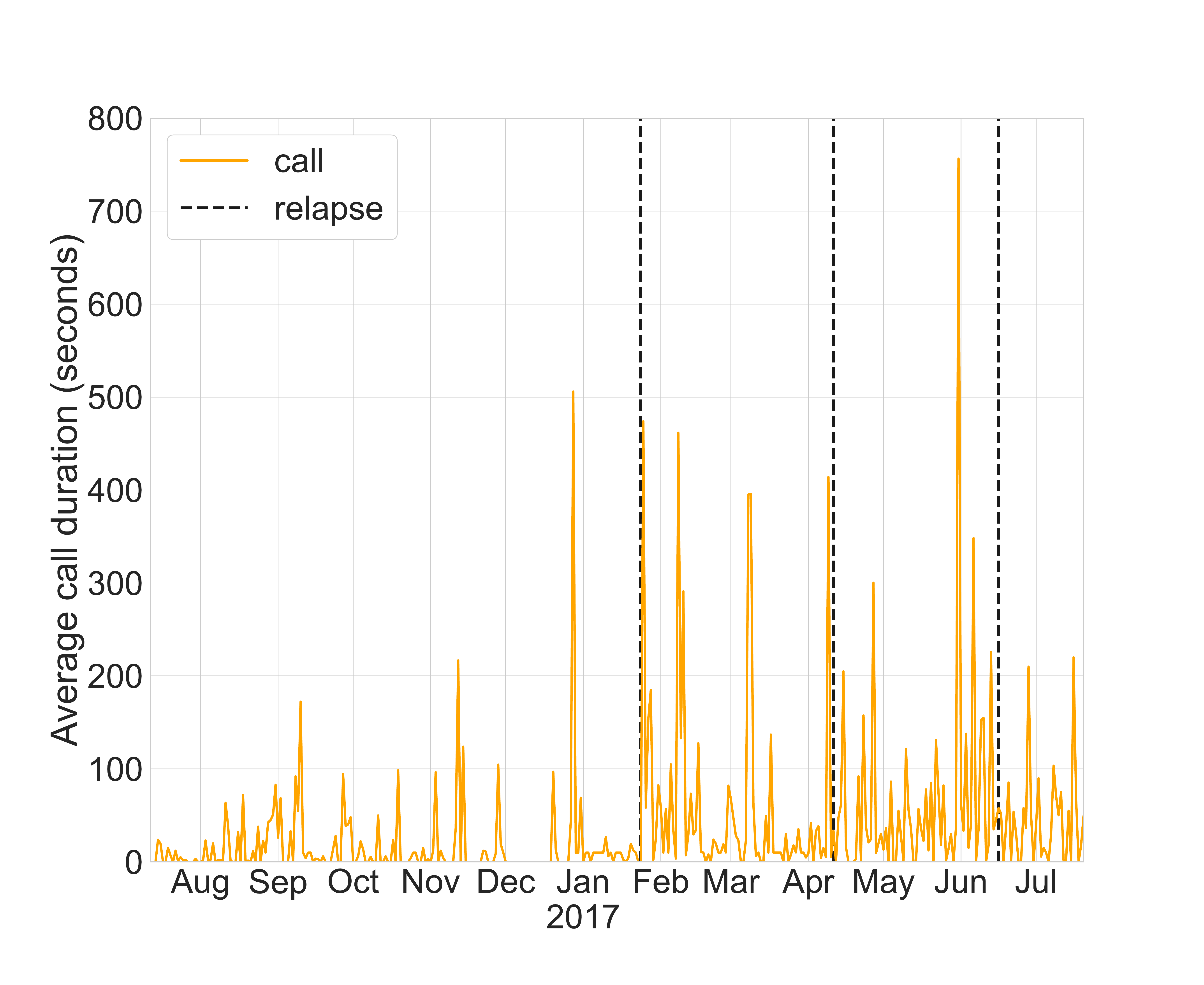}
\caption{\label{fig:call_timeseries} An example showing the call duration time-series for a patient who has had three relapses. Increased activity in call duration signal is seen near the relapse dates, though there are other similar activities in few non-relapse periods too.}
\end{figure}

The template features were obtained from 6 mobile sensor signals to characterize the behavioral patterns, and the EMA based features were extracted to further characterize the emotional state of the patient. We analyzed the classification performance obtained with the individual modalities (feature set from 6 mobile sensor signals and EMA). The demographic information is also included in the feature set for this analysis and we used the Naive Bayes based classification pipeline. The obtained result is given in Figure \ref{fig:per_modality_f2}, showing the top three modalities with the highest classification performance. Distance traveled was found to provide the best classification performance, followed by the EMA and the call duration modalities. An example of the call duration time-series for a patient who had three instances of relapse is shown in Figure \ref{fig:call_timeseries}. Increased call duration activity are seen closer to the relapse dates.

\subsubsection{Impact of feature selection and demographics features}
We implemented our relapse prediction model using mutual information based feature selection. We evaluated the performance of the Naive Bayes based classifier when no feature selection is used. Similarly, we also evaluated the classification pipeline without the demographic features, to quantify the impact of including those features. The obtained result is given in Table \ref{Table:evaluation_setting}. Both the feature selection and inclusion of demographic features were found to be advantageous for the classification performance.

\begin{table}[!htb]
\centering
\caption{\label{Table:evaluation_setting} Comparison of different evaluation setting to assess the impact of feature selection and demographics feature on classification performance.}{
\begin{tabular}{|c|c|}
\hline 
\textbf{Evaluation Setting} & \textbf{F2-score} \\ 
\hline 
All features, Feature selection, Naive Bayes model & \textbf{0.083} \\
\hline 
All features, No feature selection, Naive Bayes model & 0.036 \\
\hline
All except demographic features, Feature selection, Naive Bayes model & 0.058 \\
\hline
\end{tabular}} 
\end{table}

\section{Discussion}
\label{section:discussion}

We investigated mobile sensing based schizophrenia relapse prediction using patient-independent evaluation in this work. Our implementation of the relapse prediction model is closer to clinical deployment and builds upon the insights from the previous work in \cite{Wang2020} where mobile sensing based features were found to be associated with an upcoming relapse. We used features extracted from the daily template of the mobile sensing data, EMA, and demographics. The mobile sensing data and EMA characterized behavioral and emotional rhythmic patterns while the demographics information helped to personalize the prediction models. We obtained an F2 score of 0.083 with Naive Bayes based classifier for relapse prediction. Though this classification performance is low, it is still much higher than the random classification baseline (F2 score of 0.02). Thus, mobile sensing data has predictive value for schizophrenia relapse prediction even when employed in a patient-independent sequential prediction model, close to a clinical deployment scenario. Nonetheless, the lower F2 score obtained indicates that the relapse prediction task based on mobile sensing is difficult, and more improvements need to be done. Towards this effort, we will investigate more discriminatory features derived from the mobile sensing data (e.g. novel mobility features as presented in \cite{Canzian2015}) in future work.

We evaluated different classifiers for the relapse prediction task. The simpler Naive Bayes based classifier outperformed relatively complex Balanced Random Forest and EasyEnsemble classifiers (Table \ref{Table:results_comparison}). This could be because our dataset size is small and complex models had difficulties generalizing. We also evaluated a one-class classification (outlier detection) technique to detect relapses using Isolation Forests. Though the obtained performance was better than the random classification baseline, one-class classifier resulted in a slightly lower F2 score than those obtained with the two-class Naive Bayes classifier. This shows that supervised classification is helpful for relapse prediction, probably because the dataset size is not large enough for unsupervised approaches to automatically learn a good generalized model of the non-relapse cases.

In our work, we used daily templates composed of hourly averages of mobile sensing data to extract features characterizing behavioral patterns. This feature set was found to provide better performance when compared to the features from \cite{Wang2020} where features were computed per 6-hour epochs of the day (Table \ref{Table:feature_comparison}). A higher temporal resolution might be better to characterize finer nuances in the behavioral patterns, leading to the higher classification performance obtained. Similarly, individual signal modalities were found to provide lower classification performance (Figure \ref{fig:per_modality_f2}) compared to the classification performance obtained with multiple modalities combined together. This shows that a multi-modal assessment of behavior is important for the relapse prediction task. With a single modality, the observed behavioral pattern of an individual might be noisy and incomplete. However, with the inclusion of multiple modalities, the resulting feature dimension is also large. When the dataset is small, as in our case, feature selection is important to reduce the feature dimension (Table \ref{Table:evaluation_setting}). In our relapse prediction model, the feature selection aids for model personalization since the selected features are made dependent on the age group. Further, demographic features are also directly provided as input in the model for implicit personalization. Both of these approaches were found to be helpful for classification (Table \ref{Table:evaluation_setting}). Other feature personalization approaches need to be investigated in future work. Behavioral patterns before a relapse might manifest differently in different patients. A relapse prediction model that can adaptively personalize to the best signal modalities for a given patient, in a given period, might lead to improved classification performance.

\section{Conclusion}
\label{section:conclusion}

 Mobile sensing could be used for detecting behavioral and emotional changes associated with an oncoming schizophrenia relapse. In this work, we developed a relapse prediction model based on the features extracted from the daily template of the mobile sensing data, EMA, and demographics. Our relapse prediction model, trained in a patient-independent setting and providing a sequential relapse prediction, is closer to a clinical deployment scenario. The developed model was found to give much better performance than a random classification baseline. Thus, we conclude that the behavioral and emotional changes detected using mobile sensing have predictive value for detecting an oncoming schizophrenia relapse. The classification performance currently obtained for relapse prediction is still low and much room for improvement exists. Relapse prediction task is particularly challenging due to the limited instances of relapse incidences which makes it difficult to develop a generalized model that works across different patients. Even within the same patient, different relapse incidences might manifest differently in terms of observed behavioral and emotional changes. We will continue the investigation of optimal features and classification framework that uniquely addresses the challenges of the relapse prediction task in future work. 

%
%
%
\bibliographystyle{splncs04}
\bibliography{schizophrenia_relapse}

\begin{thebibliography}{10}
\providecommand{\url}[1]{\texttt{#1}}
\providecommand{\urlprefix}{URL }
\providecommand{\doi}[1]{https://doi.org/#1}

\bibitem{Andreasen1991}
Andreasen, N.C., Flaum, M.: Schizophrenia: The characteristic symptoms.
  Schizophrenia Bulletin  \textbf{17}(1),  27--49 (1991)

\bibitem{Barnett2018}
Barnett, I., Torous, J., Staples, P., Sandoval, L., Keshavan, M., Onnela, J.P.:
  Relapse prediction in schizophrenia through digital phenotyping: a pilot
  study. Neuropsychopharmacology  \textbf{43}(8),  1660--1666 (Jul 2018)

\bibitem{BenZeev2017}
Ben-Zeev, D., Brian, R., Wang, R., Wang, W., Campbell, A.T., Aung, M.S.H.,
  Merrill, M., Tseng, V.W.S., Choudhury, T., Hauser, M., Kane, J.M., Scherer,
  E.A.: Crosscheck: Integrating self-report, behavioral sensing, and smartphone
  use to identify digital indicators of psychotic relapse. Psychiatric
  rehabilitation journal  \textbf{40}(28368138),  266--275 (Sep 2017)

\bibitem{Birnbaum2019}
Birnbaum, M.L., Ernala, S.K., Rizvi, A.F., Arenare, E., R.~Van~Meter, A.,
  De~Choudhury, M., Kane, J.M.: Detecting relapse in youth with psychotic
  disorders utilizing patient-generated and patient-contributed digital data
  from facebook. npj Schizophrenia  \textbf{5}(1), ~17 (Oct 2019)

\bibitem{Bishop2020}
Bishop, F.M.: Relapse prediction: A meteorology-inspired mobile model. Health
  Psychology Open  (2020)

\bibitem{Buck2019}
Buck, B., Scherer, E., Brian, R., Wang, R., Wang, W., Campbell, A., Choudhury,
  T., Hauser, M., Kane, J.M., Ben-Zeev, D.: Relationships between smartphone
  social behavior and relapse in schizophrenia: A preliminary report.
  Schizophrenia research  \textbf{208},  167--172 (Jun 2019)

\bibitem{Canzian2015}
Canzian, L., Musolesi, M.: Trajectories of depression: Unobtrusive monitoring
  of depressive states by means of smartphone mobility traces analysis. In:
  Proceedings of the 2015 ACM International Joint Conference on Pervasive and
  Ubiquitous Computing. p. 1293–1304. UbiComp '15, Association for Computing
  Machinery, New York, NY, USA (2015). \doi{10.1145/2750858.2805845},
  \url{https://doi.org/10.1145/2750858.2805845}

\bibitem{Chao2004}
Chao, C., Liaw, A., Breiman, L.: Using random forest to learn imbalanced data.
  Tech. rep., University of California, Berkeley (2004)

\bibitem{Matcham2019}
Faith, M., et~al.: Remote assessment of disease and relapse in major depressive
  disorder (radar-mdd): a multi-centre prospective cohort study protocol. BMC
  Psychiatry  \textbf{19}(1), ~72 (Feb 2019)

\bibitem{Jablensky2010}
Jablensky, A.: The diagnostic concept of schizophrenia: its history, evolution,
  and future prospects. Dialogues in clinical neuroscience
  \textbf{12}(20954425),  271--287 (2010)

\bibitem{James2018}
James, S.L., et~al.: Global, regional, and national incidence, prevalence, and
  years lived with disability for 354 diseases and injuries for 195 countries
  and territories, 1990–2017: a systematic analysis for the global burden of
  disease study 2017. The Lancet  \textbf{392}(10159),  1789 -- 1858 (2018)

\bibitem{Lieberman1987}
Lieberman, J.A., Kane, J.M., Sarantakos, S., Gadaleta, D., Woerner, M., Alvir,
  J., Ramos-Lorenzi, J.: Prediction of relapse in schizophrenia. Archives of
  General Psychiatry  \textbf{44}(7),  597--603 (1987)

\bibitem{Liu2008}
{Liu}, F.T., {Ting}, K.M., {Zhou}, Z.: Isolation forest. In: 2008 Eighth IEEE
  International Conference on Data Mining. pp. 413--422 (Dec 2008)

\bibitem{Liu2009}
{Liu}, X., {Wu}, J., {Zhou}, Z.: Exploratory undersampling for class-imbalance
  learning. IEEE Transactions on Systems, Man, and Cybernetics, Part B
  (Cybernetics)  \textbf{39}(2),  539--550 (April 2009)

\bibitem{Overall1962}
Overall, J.E., Gorham, D.R.: The brief psychiatric rating scale. Psychological
  Reports  \textbf{10}(3),  799--812 (Feb 1962)

\bibitem{Shiffman2008}
Shiffman, S., Stone, A.A., Hufford, M.R.: Ecological momentary assessment.
  Annual review of clinical psychology  \textbf{4},  1--32 (2008)

\bibitem{Tseng2020}
Tseng, V.W.S., Sano, A., Ben-Zeev, D., Brian, R., Campbell, A.T., Hauser, M.,
  Kane, J.M., Scherer, E.A., Wang, R., Wang, W., Wen, H., Choudhury, T.: Using
  behavioral rhythms and multi-task learning to predict fine-grained symptoms
  of schizophrenia. Scientific Reports  \textbf{10}(1),  15100 (Sep 2020).
  \doi{10.1038/s41598-020-71689-1},
  \url{https://doi.org/10.1038/s41598-020-71689-1}

\bibitem{Wang2020}
{Wang}, R., {Wang}, W., {Ben-Zeev}, D., {Brian}, R., {Campbell}, A.,
  {Choudhury}, T., {Hauser}, M., {Kane}, J., {Obuchi}, M., {Scherer}, E.,
  {Walsh}, M.: Methods for predicting relapse episodes in schizophrenia using
  mobile phone sensing. In: 2020 IEEE International Conference on Pervasive
  Computing and Communications (PerCom) (2020)

\bibitem{Wang2018}
Wang, R.: Mental Health Sensing Using Mobile Phones. Ph.D. thesis, Dartmouth
  College (2018)

\bibitem{Wang2016}
Wang, R., Aung, M.S.H., Abdullah, S., Brian, R., Campbell, A.T., Choudhury, T.,
  Hauser, M., Kane, J., Merrill, M., Scherer, E.A., Tseng, V.W.S., Ben-Zeev,
  D.: Crosscheck: Toward passive sensing and detection of mental health changes
  in people with schizophrenia. In: Proceedings of the 2016 ACM International
  Joint Conference on Pervasive and Ubiquitous Computing. p. 886–897. UbiComp
  ’16, Association for Computing Machinery, New York, NY, USA (2016)

\bibitem{Wang2017}
Wang, R., Wang, W., Aung, M.S.H., Ben-Zeev, D., Brian, R., Campbell, A.T.,
  Choudhury, T., Hauser, M., Kane, J., Scherer, E.A., Walsh, M.: Predicting
  symptom trajectories of schizophrenia using mobile sensing. Proc. ACM
  Interact. Mob. Wearable Ubiquitous Technol.  \textbf{1}(3) (Sep 2017)

\bibitem{yang2018}
Yang, Z., Nguyen, L., Jin, F.: Predicting opioid relapse using social media
  data (2018), \url{https://arxiv.org/pdf/1811.12169.pdf}

\end{thebibliography}
\end{document}